\documentclass[10pt]{iopart}

\usepackage{iopams}
\usepackage{bm} 

\newtheorem{theorem}{Theorem}

\newtheorem{corollary}{Corollary}
\newtheorem{definition}{Definition}

\begin{document}

\title{Controllability of open quantum systems with Kraus-map dynamics}
\author{Rong Wu, Alexander Pechen, Constantin Brif and Herschel Rabitz}
\address{Department of Chemistry, Princeton University,
Princeton, New Jersey 08544}
\eads{\mailto{apechen@princeton.edu},
\mailto{cbrif@princeton.edu},
\mailto{hrabitz@princeton.edu}}

\begin{abstract}
This paper presents a constructive proof of complete
kinematic state controllability of finite-dimensional open
quantum systems whose dynamics are represented by Kraus
maps. For any pair of states (pure or mixed) on the
Hilbert space of the system, we explicitly show how to
construct a Kraus map that transforms one state into
another. Moreover, we prove by construction the existence
of a Kraus map that transforms \emph{all} initial states
into a predefined target state (such a process may be
used, for example, in quantum information dilution). Thus,
in sharp contrast to unitary control, Kraus-map dynamics
allows for the  design of controls which are robust to
variations in the initial state of the system. The
capabilities of non-unitary control for population
transfer between pure states illustrated for an example of
a two-level system by constructing a family of non-unitary
Kraus maps to transform one pure state into another. The
problem of dynamic state controllability of open quantum
systems (i.e., controllability of state-to-state
transformations, given a set of available dynamical
resources such as coherent controls, incoherent
interactions with the environment, and measurements) is
also discussed.
\end{abstract}

\submitto{\JPA}

\section{Introduction}

Coherent control of quantum systems is a rapidly developing area of
research with applications to numerous physical and chemical
problems~\cite{RVM00,RaWa03,RZ00,SB03,DL04}. The general goal of
quantum control is to manipulate the dynamics of a quantum system in a
desired way by applying suitable external control fields, typically,
optimally shaped pulses of a coherent radiation field. Much
theoretical work \cite{PeDaRa,WRD93,RHR} has been devoted to coherent
control of closed quantum systems with unitary dynamics.  However,
realistic physical situations entail control of open quantum systems
whose dynamics is non-unitary due to interactions with the
environment. Research on various aspects of control of open quantum
systems has appeared in recent years
\cite{Ll00,LV02,SS02,Alt,JP05,Ro05,MM03,MC05,Zhang06,RDA06,Pechen,Shuang},
motivated by many applications including quantum computing
\cite{NiCh00,Preskill,Tarasov,Grace06}, laser cooling
\cite{Wineland79,TKB99,Schirmer01,STK04}, quantum reservoir
engineering \cite{PCZ96}, management of decoherence
\cite{Brif01,ZhRa03,Vi04,Vi05,KhLi05,Grig05,SB06,Grace06b,%
Fel05,Mort06,Brand06}, chemical reactions and energy
transfer in molecules \cite{Assion,Umeda,Levis,Rom,Herek}.

A coherent control field acts on the system through the Hamiltonian
part of its dynamics. A qualitatively different approach relies on
using specially tailored environments, which affect the system via
non-unitary evolution, with controls applied through the dissipative
part of the dynamics~\cite{ice}. In this approach, a suitably
optimized non-equilibrium distribution function of an environment
(e.g., an electron, atom, or molecular gas, or a solvent) is employed
as a control instrument to achieve the desired objective. This type of
incoherent control by the environment (ICE) may be combined with
optimally tailored coherent fields to allow for simultaneous control
through both the Hamiltonian and dissipative parts of the system
dynamics.

One of the fundamental issues of quantum control is assessing the
system's controllability. A quantum system is controllable in a set of
configurations, $\mathcal{S} = \{ \lambda \}$, if for any pair of
configurations $\lambda_1, \lambda_2 \in{\cal S}$ there exists a
time-dependent control, $c(t)$, that can drive the system from the
initial configuration $\lambda_1$ to the final configuration
$\lambda_2$ in a finite time $T$. Here, the notion of configuration
means either the state of the system $\rho$, the expectation value of
an observable $\mathrm{Tr}(\rho O)$, the evolution operator $U(t)$, or
the Kraus map $\Phi$, depending on the specific control problem.
Controllability of closed quantum systems with unitary dynamics has
been extensively studied
\cite{HTC83,Rama95,TuRa,SFS01,Alt02,GSLK98,SL01,SSL02,ADA03,SPS05,%
CLT03,WTL06}. We will briefly review some of these relevant results in
section~\ref{sec:problem}.

Unitary dynamics can achieve control only within sets of states
exhibiting the same density-matrix spectrum, and a unitary
transformation cannot connect two quantum states of different
purity. Non-unitary evolution of open quantum systems is able to lift
this restriction and transform pure states into mixed ones and vice
versa (a familiar example is the cooling of a thermalized quantum
system, which requires coupling to a reservoir). However, the
important question of controllability of open quantum systems is not
yet fully addressed, although some aspects of this problem have been
considered. In particular, controllability of a quantum system
undergoing non-unitary evolution and controlled by a coherent field,
that acts only through the Hamiltonian part of the dynamics, has been
discussed in a number of works~\cite{SS02,Alt,JP05}. Another related
research direction concerns supplementing unitary coherent controls by
measurements \cite{LV02,MM03,MC05,Zhang06,Pechen,Shuang}.

In this paper, we take a different perspective by
considering the problem of kinematic state controllability
(KSC) of open quantum systems whose dynamics are
represented by Kraus maps \cite{Kraus83,AlLe87}.
Specifically, we prove the existence of a Kraus map that
can move a finite-dimensional open quantum system from any
initial state $\rho_{\rm in}$ to any final target state
$\rho_{\rm f}$. This establishes complete KSC of
finite-dimensional open quantum systems with Kraus-map
dynamics, in contrast to restricted KSC of closed quantum
systems where unitary dynamics can connect only states
with the same density-matrix spectrum.

The constructed Kraus map transforms \emph{all} initial
states into a predefined final target state $\rho_{\rm
f}$. Such Kraus transformations can be used, for example,
in the context of quantum information
dilution~\cite{Ziman02,Roa06} to realize a mapping of an
unknown (mixed or pure) quantum state into a given target
state. The existence of such all-to-one maps is a
significant distinction between non-unitary evolution and
unitary evolution, since in the latter case the evolution
operator and the corresponding coherent control field
always depend on both the initial and target states of the
system. Therefore, extending the controls to include
appropriate non-unitary dynamics allows for solving the
problem of achieving control operations which are robust
to variations in the initial state of the system.
Robustness to variations in the initial state is
understood here as the ability to use a single control
(i.e., a single Kraus map) to transfer all initial states
into a predefined, mixed or pure, final target state. The
possibility of using a single Kraus map to transform all
initial states into a given final state is a property
stronger than transitivity of the set of Kraus maps on the
set of density matrices in a finite-dimensional Hilbert
space. This property does not have an analogue for unitary
transformations. While the set of unitary operators acts
transitively on the set of unit-norm vectors, no single
unitary transformation can map all pure states into a
given final pure state.

In practice, the unitary or non-unitary dynamics of the
system is guided by a set of available controls. Possible
controls include pulses of coherent electromagnetic
radiation, incoherent environments (e.g., electron, atom,
or molecular gases, or a solvent) with tunable
non-equilibrium distribution functions, disturbances
induced by quantum measurements, etc. The ability to make
transformations between the states of the system, using
the available set of controls, is referred to as dynamic
state controllability (DSC). For a specific problem, DSC
is determined by the particular dependence of the Kraus
operators on the controls. In this paper, we discuss some
general properties of DSC which do not require knowledge
of this dependence.

Several important problems still remain open, including a
study of DSC for specific quantum systems, taking into
account the available laboratory control tools, and an
analysis of robustness of the dynamical control to
imperfections and environmental couplings during the
evolution. Some attempts in this direction exist,
including, for example, an analysis of coherent control of
non-dispersive wave packets~\cite{Buchleitner02}, although
the problem remains generally open for a future study.
However, first it would be desirable to lay the ground for
these system-dependent studies by performing a general
analysis of KSC and DSC, which can reveal the highest
degree of control attained by general physically allowable
dynamics. The existence of quantum controls which are
robust to variations in the initial states, established in
the present work, should facilitate an exploration of
non-unitary control tools for specific systems.

Experimental studies of environmentally-induced
decoherence in open quantum systems explored various
physical processes, including the loss of spatial
coherence of an atomic wave-function due to spontaneous
emission~\cite{Pfau94}, decoherence of motional
superposition states of a trapped atom coupled to
engineered reservoirs~\cite{Myatt00}, the loss of spatial
coherence in matter-wave interferometry with fullerenes
caused by collisions with an environmental gas
\cite{Horn03} or by thermal emission of radiation
\cite{Hack04}, and decoherence in networks of spin qubits
due to pairwise dipolar interactions between the spins
\cite{MST06}. Control of decoherence was experimentally
explored in several systems, including photon pairs
generated from atomic ensembles \cite{Fel05}, nuclear spin
qubits in fullerenes \cite{Mort06}, and vibrational
wave-packets in diatomic molecules \cite{Brand06}.
Suppression of decoherence for quantum computing using
environment induced quantum Zeno effect was
suggested~\cite{Beige00}. Control of decoherence of
motional quantum states caused by scattering events was
proposed in~\cite{Va05}. The feasibility of decoherence
suppression via manipulations of quantum states is
evidently related to a more general problem of state
controllability. For example, the controllability analysis
can be applied to examine the feasibility of transforming
an arbitrary initial state into a state in a
weak-decoherence subspace. Considering the rapidly growing
interest in experimental management of decoherence
\cite{Fel05,Mort06,Brand06}, our theoretical analysis
hopefully will stimulate further laboratory studies of
practical aspects of state controllability in open quantum
systems.

The paper is organized as follows. In
section~\ref{sec:problem} we formulate definitions of
controllability for closed and open quantum systems. The
general proof of open-system KSC is presented in
section~\ref{sec:proof}. The capabilities of non-unitary
control for population transfer between pure states are
illustrated in section~\ref{sec:examples} for two-level
quantum systems. In section~\ref{sec:DSC} we also discuss
general conditions for DSC of open quantum systems
undergoing Kraus-map evolution.

\section{Formulation of the controllability analysis}
\label{sec:problem}

\subsection{Definitions of controllability for closed quantum
systems}
\label{sec:closed}

A discussion of different notions of controllability in
finite-dimensional closed quantum systems with unitary dynamics is
available~\cite{SSL02,ADA03}. Here, we briefly review some basic
definitions and results as background for consideration of the
open-system controllability analysis to follow.

For unitary dynamics, KSC is defined as follows:
\begin{definition} \label{def:KSC}
A closed quantum system with unitary dynamics is kinematically
controllable in a set $\mathcal{S}_{\mathrm{K}}$ of states if for any
pair of states $\rho_1 \in \mathcal{S}_{\mathrm{K}}$ and $\rho_2 \in
\mathcal{S}_{\mathrm{K}}$ there exists a unitary operator $U$, such
that $\rho_2 = U \rho_1 U^{\dagger}$.
\end{definition}
Any two quantum states that belong to the same kinematically
controllable set $\mathcal{S}_{\mathrm{K}}$ are called kinematically
equivalent. It is straightforward to see \cite{SSL02} that two states
$\rho_1$ and $\rho_2$ of a closed quantum system are kinematically
equivalent if and only if they have the same eigenvalues. Therefore,
all quantum states that belong to the same kinematically controllable
set have the same density-matrix eigenvalues, the same von Neumann
entropy, and the same purity. For example, all pure states belong to
the same kinematically controllable set. However, any pure state is
not kinematically equivalent to any mixed state. For a closed quantum
system all states on the system's Hilbert space are separated into
unconnected sets of kinematically equivalent states.

The dynamics of a closed quantum system is governed by the
Schr\"odinger equation:
\begin{equation} \label{eq:Schr}
\rmi \hbar \frac{\rmd U(t)}{\rmd t} = H U(t), \qquad U(0) = I .
\end{equation}
Here, $H$ is the Hamiltonian, $U(t)$ is the evolution operator, and
$I$ is the identity operator.  Assuming that the Hamiltonian $H$ is a
functional of a set of time-dependent controls $H = H[c_1(t), \ldots ,
c_k(t)]$, DSC for unitary evolution is defined as follows:
\begin{definition} \label{def:DSC}
A closed quantum system with unitary evolution is dynamically
controllable in a set $\mathcal{S}_{\mathrm{D}}$ of states if for any
pair of states $\rho_1 \in \mathcal{S}_{\mathrm{D}}$ and $\rho_2 \in
\mathcal{S}_{\mathrm{D}}$ there exist a finite time $T$ and a set of
controls $\{ c_1(t), \ldots , c_k(t) \}$, such that the solution
$U(T)$ of the Schr\"odinger equation (\ref{eq:Schr}) transforms
$\rho_1$ into $\rho_2$: $\rho_2 = U(T) \rho_1 U^{\dagger}(T)$.
\end{definition}
Since unitary dynamics can be controlled only within the set of
kinematically equivalent states, a dynamically controllable set of
states $\mathcal{S}_{\mathrm{D}}$ is always a subset of the
corresponding kinematically controllable set
$\mathcal{S}_{\mathrm{K}}$. If the dynamically controllable set of
pure states coincides with its kinematically controllable counterpart
(i.e., the set of all pure states), the closed quantum system is
called pure-state controllable. If all dynamically controllable sets
of states coincide with their kinematically controllable counterparts,
the system is called density-matrix controllable.

It is possible to define controllability of a closed quantum system
not only in a set of states, but also in a set of evolution operators
$U(t)$. The corresponding property, called evolution-operator
controllability (EOC), is defined as follows:
\begin{definition} \label{def:EOC}
A closed quantum system with unitary dynamics is evolution-operator
controllable if for any unitary operator $V$ there exists a finite
time $T$ and a set of controls $\{ c_1(t), \ldots , c_k(t) \}$, such
that $V = U(T)$, where $U(T)$ is the solution of the Schr\"odinger
equation (\ref{eq:Schr}) with $H = H[c_1(t), \ldots , c_k(t)]$.
\end{definition}
For an $N$-level closed system, a necessary and sufficient condition
for EOC is \cite{SSL02,ADA03} that the dynamical Lie group
$\mathfrak{G}$ of the system be U($N$) [or SU($N$) for a traceless
Hamiltonian, which differs from the original one just by a physically
irrelevant shift in the energy]. It can be also shown
\cite{SSL02,ADA03} that EOC is equivalent to density-matrix
controllability, while the condition for pure-state controllability is
weaker.

\subsection{Definition of KSC for open quantum systems}

The state of an open quantum system is represented by the
reduced density matrix $\rho =
\mathrm{Tr}_{\mathrm{E}}(\rho_{\mathrm{tot}})$, where
$\rho_{\mathrm{tot}}$ is the density matrix of the system
and environment taken together, and
$\mathrm{Tr}_{\mathrm{E}}$ denotes the trace over the
environment degrees of freedom. The dynamics of open
quantum systems is governed by various master equations
(see, e.g.,~\cite{Spohn} on master equations for a system
weakly interacting with the environment and recent works
on master equations for collisional decoherence with a
strong interaction~\cite{APV,Hornberger,Adler,Halliwell}).
If the system and environment are initially uncorrelated,
the time evolution of the system in the kinematic picture
can be described by a completely positive,
trace-preserving linear map.

Let $\mathcal{H}$ be the Hilbert space of the system and
$\mathcal{T}(\mathcal{H})$ be the space of trace-class operators on
$\mathcal{H}$. For example, for an $N$-level quantum system,
$\mathcal{H} = \mathbb{C}^N$ is the space of complex vectors of length
$N$ and $\mathcal{T}(\mathcal{H}) = \mathcal{M}_N$ is the space of $N
\times N$ complex matrices. The set of density matrices (i.e., the set
of positive operators on ${\cal H}$ with trace one) is denoted as
$\mathcal{D}(\mathcal{H})$ [clearly, $\mathcal{D}(\mathcal{H}) \subset
  \mathcal{T}(\mathcal{H})$].
\begin{definition}
A linear map $\Phi : \mathcal{T}(\mathcal{H}) \to
\mathcal{T}(\mathcal{H})$ is called completely positive if
the map $\Phi \otimes I_l : \mathcal{T}(\mathcal{H})
\otimes \mathcal{M}_l \to \mathcal{T}(\mathcal{H}) \otimes
\mathcal{M}_l$ (where $I_l$ is the identity map in
$\mathcal{M}_l$) is positive for any $l \in \mathbb{N}$.
The map $\Phi$ is called trace preserving if for any $\rho
\in \mathcal{T}(\mathcal{H})$, $\mathrm{Tr}\, (\Phi[\rho])
= \mathrm{Tr}\, (\rho)$.
\end{definition}

Any completely positive, trace-preserving map has the Kraus
operator-sum representation~\cite{Kraus83,AlLe87,Choi}:
\begin{equation}
\label{eq:Kraus-map} \Phi[\rho] = \sum_{i=1}^{n} K_i \rho
K_i^{\dagger} ,
\end{equation}
where the Kraus operators $K_i$ satisfy the condition
\begin{equation}
\label{eq:k_condi} \sum_{i=1}^{n} K_i^{\dagger} K_i = I .
\end{equation}
Here, $n \in \mathbb{N}$ is the number of the Kraus operators $K_i$
and $I$ is the identity operator on $\mathcal{H}$. The condition
(\ref{eq:k_condi}) ensures the preservation of the trace:
$\mathrm{Tr}\, (\Phi[\rho]) = \mathrm{Tr}\, (\rho)$. In this paper, we
refer to completely positive, trace-preserving maps simply as Kraus
maps. Unitary transformations of the system states form a particular
subset of Kraus maps corresponding to $n = 1$. Note that a
composition of any two Kraus maps $\Phi_1$ and $\Phi_2$ is another
Kraus map:
\begin{equation}
\label{eq:semigroup} \Phi_2 [ \Phi_1 [\rho] ] \equiv
(\Phi_2 \circ \Phi_1) [\rho] = \Phi_3 [\rho] .
\end{equation}

It is well known that any Kraus map $\Phi$ has infinitely many
different Kraus operator-sum representations of the form
(\ref{eq:Kraus-map}). Let $\{ K_1, \ldots, K_n \}$ be a set of Kraus
operators representing $\Phi$. For $m \geq n$, consider an $m \times
n$ matrix $W$ with elements $w_{i j}$, such that $W^{\dagger} W =
I_n$. Define a new set of Kraus operators:
\begin{equation}\label{eq1}
\tilde{K}_i = \sum_{j=1}^n w_{i j} K_j , \qquad i = 1, \ldots, m .
\end{equation}
Then for any $\rho\in\mathcal{T}(\mathcal{H})$, one has
\begin{equation}
\sum_{i=1}^m \tilde{K}_i \rho \tilde{K}_i^{\dagger} =
\sum_{i=1}^n K_i \rho K^{\dagger}_i ,
\end{equation}
i.e., both sets of Kraus operators, $\{ K_1, \ldots, K_n \}$ and $\{
\tilde{K}_1, \ldots, \tilde{K}_m \}$, represent the same Kraus map
$\Phi$. Moreover, if two different sets of Kraus operators represent
the same Kraus map, then they are necessarily related by (\ref{eq1})
with a matrix $W$ such that $W^{\dagger} W = I$. Any Kraus map for an
$N$-level quantum system can be represented by a set of $n \leq N^2$
Kraus operators \cite{Choi}. That is, if the map is represented by a
set of $n > N^2$ Kraus operators, there always exists another
representation with not more than $N^2$ operators.

In the context of the system coupled to the environment, complete
positivity of a map $\Phi$ means that for any admissible evolution of
the system density matrix, $\Phi[\rho]$, the initial density matrix
$\rho_{\mathrm{tot}}(0)$ of the system and environment taken together
will evolve into another density matrix \cite{Preskill}. If the system
and environment are initially correlated, the map describing the
evolution of the system density matrix will not be always completely
positive and the Kraus operator-sum representation
(\ref{eq:Kraus-map}) will not be always valid \cite{HKO03,JSS04}.

The notions of closed-system controllability presented in
section~\ref{sec:closed} need to be modified for open quantum systems
which allow non-unitary dynamics. For an open quantum system with
Kraus-map dynamics, KSC is defined as follows:
\begin{definition} \label{def:KSC-Kraus}
An open quantum system with the Kraus-map dynamics of the form
(\ref{eq:Kraus-map}) is kinematically controllable in a set
$\mathcal{S}_{\mathrm{K}}$ of states if for any pair of states
$\rho_1 \in \mathcal{S}_{\mathrm{K}}$ and
$\rho_2 \in \mathcal{S}_{\mathrm{K}}$ there exists a Kraus map $\Phi$,
such that $\rho_2 = \Phi[ \rho_1 ]$.
\end{definition}
In the next section we will prove that KSC for a finite-dimensional
open system with Kraus-map dynamics is \emph{complete}, i.e., that the
system is kinematically controllable in the set
$\mathcal{S}_{\mathrm{K}} = \mathcal{D}(\mathcal{H})$ of \emph{all}
density operators $\rho$ on the Hilbert space $\mathcal{H}$. The
problem of DSC for the Kraus-map evolution will be discussed in
section~\ref{sec:DSC}.

\section{Proof of complete KSC for open quantum systems with
Kraus-map dynamics}
\label{sec:proof}

The proof of complete KSC for open quantum systems with Kraus-map
dynamics is given here by construction. We start by reformulating the
known result that Kraus maps can transform all states of a
finite-dimensional open quantum system into a given pure state. Then
we prove that open quantum systems with Kraus map dynamics are
completely kinematically controllable. Moreover, we prove by
construction a stronger result --- the existence of Kraus maps which
transform all states of an open quantum system into an arbitrary (not
necessarily pure) target state.  We also construct Kraus maps which
transform a given pure state into an arbitrary (mixed or pure) target
state.

It is known (see, for example, section~8.3.1 of \cite{NiCh00})
that Kraus maps can transform mixed states into pure ones. We now
reformulate this result in a general way.
\begin{theorem} \label{lemma1}
For any pure state $\rho_{\mathrm{p}} = |\psi
\rangle\langle \psi|$ on the Hilbert space $\mathcal{H}$
of a finite-dimensional open quantum system, there exists
a Kraus map $\Phi_{\mathrm{atp}}$, such that
$\Phi_{\mathrm{atp}} [\rho] = \rho_{\mathrm{p}}$ for
\emph{all} states $\rho$ on
$\mathcal{H}$\footnote[1]{Abbreviation atp of {\it all to
pure} is used to indicate that the Kraus map $\Phi_{\rm
atp}$ transforms all states into a pure state.}.
\end{theorem}
\emph{Proof}: Define the operators $K_i^{(\mathrm{atp})} = | \psi
\rangle \langle \chi_i |$ for $i = 1, \ldots, N$, where $N$ is the
dimension of the system and $\{ \chi_i \}$ is an arbitrary orthonormal
basis in $\mathcal{H}$. The operators $K_i^{(\mathrm{atp})}$ satisfy
the normalization condition (\ref{eq:k_condi}) and define the Kraus map
\begin{equation} \label{eq:Phi-atp}
\Phi_{\mathrm{atp}} [\rho] = \sum_{i=1}^N
K_i^{(\mathrm{atp})} \rho K_i^{(\mathrm{atp}) \dagger} .
\end{equation}
This map transforms all states into the pure state
$\rho_{\mathrm{p}}$ since for any density matrix $\rho$
\begin{equation} \label{eq:Phi-atp-action}
\Phi_{\mathrm{atp}} [\rho]
= \sum_{i=1}^N |\psi \rangle \langle \chi_i |
\rho |\chi_i \rangle\langle \psi |
= ( \mathrm{Tr}\,\rho ) |\psi\rangle\langle \psi|
= \rho_{\mathrm{p}} .
\end{equation}
This completes the proof.

The choice of the orthonormal basis $\{ \chi_i \}$ in the proof above
is completely arbitrary. Different bases $\{ \chi_i \}$ determine
different sets of Kraus operators $\{ K_i^{(\mathrm{atp})} \}$ which
are related to each other by (\ref{eq1}) and all represent the same
Kraus map $\Phi_{\mathrm{atp}}$. The map $\Phi_{\mathrm{atp}}$
transforms all initial states $\rho$ into the same final state
$\rho_{\mathrm{p}}$, i.e., it is an all-to-one map.

Next, we generalize the result of Theorem~\ref{lemma1} to the case of
an arbitrary (not necessarily pure) target state.
\begin{theorem} \label{t1}
For any state $\rho_{\mathrm{f}}$ on the Hilbert space $\mathcal{H}$
of a finite-dimensional open quantum system, there exists a Kraus map
$\Phi$ such that $\Phi[\rho]=\rho_{\mathrm{f}}$ for \emph{all} states
$\rho$ on $\mathcal{H}$.
\end{theorem}
\emph{Proof}: Let the spectral decomposition of the final
state $\rho_{\mathrm{f}}$ be
\begin{equation} \label{eq:spectr}
\rho_{\mathrm{f}} = \sum_{i=1}^N p_i |\phi_i \rangle\langle \phi_i | ,
\end{equation}
where $p_i$ is the probability to find the system in the state
$|\phi_i\rangle$ ($p_i \geq 0$ and $\sum_{i=1}^N p_i = 1$). In
particular, a pure state $\rho_{\mathrm{f}}=|\phi \rangle\langle \phi
|$ has only one non-zero eigenvalue, $p_1 = 1$. For an arbitrary
orthonormal basis $\{\chi_j\}$ in $\mathcal{H}$, define the operators
\begin{equation} \label{eq:K-tot}
K_{i j}= \sqrt{p_i}\, | \phi_i \rangle\langle \chi_j | ,
\qquad i , j = 1, \ldots, N .
\end{equation}
The operators $K_{i j}$ satisfy the normalization condition
(\ref{eq:k_condi}) and define the Kraus map:
\begin{equation} \label{eq:Phi-tot-2}
\Phi[\rho] = \sum_{i,j=1}^N K_{i j} \rho K_{i j}^{\dagger}.
\end{equation}
The map $\Phi$ acts on any state $\rho$ on $\mathcal{H}$ as
\begin{equation}
\Phi[\rho] = \sum_{i,j=1}^N p_i | \phi_i \rangle\langle \chi_j | \rho
|\chi_j \rangle\langle \phi_i | = ( \mathrm{Tr}\, \rho) \sum_{i=1}^N
p_i | \phi_i \rangle\langle \phi_i | = \rho_{\mathrm{f}} .
\label{eq:Phi-action-2}
\end{equation}
This completes the proof.

Complete KSC for open quantum systems with Kraus-map dynamics directly
follows from Theorem~\ref{t1} and can be expressed in the form of
corollaries.
\begin{corollary} \label{t1w}
For any pair of states $\rho_1$ and $\rho_2$ on the Hilbert space
$\mathcal{H}$ of a finite-dimensional open quantum system, there
exists a Kraus map $\Phi$ such that $\Phi[\rho_1] = \rho_2$.
\end{corollary}
\begin{corollary} \label{c1}
A finite-dimensional open quantum system with Kraus-map dynamics is
kinematically controllable in the set $\mathcal{S}_{\mathrm{K}} =
\mathcal{D}(\mathcal{H})$ of \emph{all} density operators on
$\mathcal{H}$.
\end{corollary}

Since the choice of the orthonormal basis $\{ \chi_j \}$ is completely
arbitrary, there exist infinitely many sets of Kraus operators of the
form (\ref{eq:K-tot}) (corresponding to different choices of $\{
\chi_j \}$), all of which represent the same Kraus map $\Phi$ of
(\ref{eq:Phi-tot-2}). The Kraus map $\Phi$ transforms all initial
states into the final target state $\rho_{\mathrm{f}}$, i.e., it is an
all-to-one map.

Given a pair of states $\rho_{\rm in}$ and $\rho_{\rm f}$
on the Hilbert space of a finite-dimensional open quantum
system, there exist many different Kraus maps transforming
$\rho_{\rm in}$ into $\rho_{\rm f}$. Consider, for
example, a pure initial state $\rho_{\mathrm{in}} = |\psi
\rangle\langle \psi|$ and an arbitrary (mixed or pure)
final state $\rho_{\mathrm{f}} = \sum_{i=1}^N p_i
|\phi_i\rangle\langle\phi_i|$. Let $\mathcal{U} = \{ U_i
\}_{i=1}^N$ be a set of unitary operators such that $U_i
|\psi\rangle = |\phi_i\rangle$. Define the operators
\begin{equation} \label{eq:K-pta}
K_i^{(\mathrm{pta})} = \sqrt{p_i}\, U_i , \qquad i = 1,
\ldots, N ,
\end{equation}
which satisfy the normalization condition
(\ref{eq:k_condi}) and determine the Kraus
map\footnote[2]{Abbreviation pta of {\it pure to any} is
used to indicate that the corresponding Kraus map
transforms a specific pure state into a given mixed or
pure state.}
\begin{equation} \label{eq:Phi-pta}
\Phi_{\mathrm{pta}} [\rho] = \sum_{i=1}^N
K_i^{(\mathrm{pta})} \rho K_i^{(\mathrm{pta}) \dagger} .
\end{equation}
The Kraus map $\Phi_{\mathrm{pta}}$ transforms the pure
state $\rho_{\mathrm{in}}$ into the state
$\rho_{\mathrm{f}}$:
\begin{equation} \label{eq:Phi-pta-action}
\Phi_{\mathrm{pta}} [ \rho_{\mathrm{in}} ] = \sum_{i=1}^N
p_i U_i |\psi \rangle\langle \psi| U_i^{\dagger} =
\sum_{i=1}^N p_i |\phi_i \rangle\langle \phi_i | =
\rho_{\mathrm{f}} .
\end{equation}
The choice of a set $\mathcal{U} =\{ U_i \}$ of unitary
operators with the property $U_i |\psi\rangle =
|\phi_i\rangle$ is not unique. Given any such a set ${\cal
U}$, the corresponding Kraus map $\Phi_{\rm pta}$ will be
denoted as $\Phi^{({\cal U})}_{\rm pta}$. Different sets
$\mathcal{U} =\{ U_i \}$ and $\tilde{\mathcal{U}} = \{
\tilde{U}_i \}$, where $U_i |\psi\rangle = \tilde{U}_i
|\psi\rangle = |\phi_i\rangle$, can produce different
Kraus maps $\Phi^{(\mathcal{U})}_{\mathrm{pta}}$ and
${\Phi}^{(\tilde{\mathcal{U}})}_{\mathrm{pta}}$,
respectively. All these maps satisfy
(\ref{eq:Phi-pta-action}) when they act on the particular
state $\rho_{\mathrm{in}} = |\psi\rangle\langle \psi|$,
but in general $\Phi^{(\mathcal{U})}_{\mathrm{pta}}[\rho]
\neq {\Phi}^{(\tilde{\mathcal{U}})}_{\mathrm{pta}}[\rho]$
if $\rho \neq \rho_{\mathrm{in}}$. Every
$\Phi_{\mathrm{pta}}$ is a one-to-one map, i.e., in
general $\Phi_{\mathrm{pta}} [\rho] \neq
\Phi_{\mathrm{pta}} [\rho']$ if $\rho \neq \rho'$.

For a given final state $\rho_{\mathrm{f}}$ of the form
(\ref{eq:spectr}), we find that the Kraus operators $K_{i
j}$ of (\ref{eq:K-tot}) can be constructed as
\begin{equation}
K_{i j} = K_i^{(\mathrm{pta})} K_j^{(\mathrm{atp})} =
\sqrt{p_i}\, U_i |\psi\rangle \langle \chi_j | =
\sqrt{p_i}\, |\phi_i \rangle \langle \chi_j | ,
\end{equation}
for arbitrary choices of $|\psi\rangle$, $\{ U_i \}$, and
$\{ \chi_j \}$. Therefore, the Kraus map $\Phi$ of
(\ref{eq:Phi-tot-2}), which transforms all states into a
given final state $\rho_{\mathrm{f}}$, can be constructed
as the composition of the maps $\Phi_{\mathrm{atp}}$ and
$\Phi_{\mathrm{pta}}$,
\begin{equation} \label{eq:Phi-tot}
\Phi = \Phi_{\mathrm{pta}} \circ \Phi_{\mathrm{atp}} .
\end{equation}
Indeed, using (\ref{eq:Phi-atp-action}) and
(\ref{eq:Phi-pta-action}), we obtain
\begin{equation} \label{eq:Phi-action}
\Phi [ \rho] = \Phi_{\mathrm{pta}} [ \Phi_{\mathrm{atp}} [
\rho] ] = \Phi_{\mathrm{pta}} [ |\psi\rangle\langle\psi| ]
= \rho_{\mathrm{f}} ,
\end{equation}
for all states $\rho$ on $\mathcal{H}$.

Different constructions [equations (\ref{eq:Phi-tot}) and
(\ref{eq:Phi-tot-2}), respectively] used for obtaining the
same Kraus map $\Phi$ indicate the possibility of steering
the system to the target state via different control
pathways. The construction of (\ref{eq:Phi-tot}) can be
interpreted as a two-step process in which the system is
first driven to a specific pure state (which is not
necessarily the ground state) and subsequently transformed
from this pure state into the target state (which can be
either pure or mixed). The construction of
(\ref{eq:Phi-tot-2}) describes a transformation to the
target state with no intermediate pure states involved. An
example of such a process is the evolution of a system
coupled to a thermal reservoir kept at the inverse
temperature $\beta$. In this case, under some general
conditions on the system-environment interaction, all
initial system states will eventually evolve into the same
thermal state $\rho=e^{-\beta H_0}/\mathrm{Tr} ( e^{-\beta
H_0} )$, where $H_0$ is the free system Hamiltonian. At
that, a mixed initial state will always stay mixed during
this type of evolution.

\section{Non-unitary transformations between pure states}
\label{sec:examples}

In coherent control, transitions between pure states are achieved via
unitary transformations. Unitary dynamics correspond to keeping only
one term in the Kraus operator-sum representation
(\ref{eq:Kraus-map}). Here we show, using as an example a two-level
open quantum system, that a multitude of \emph{non-unitary} Kraus maps
can be used for transforming one pure state into another.

Let the initial and final pure states be the ground $|0\rangle$ and
excited $|1\rangle$ states of a two-level system, with density
matrices $\rho_{\mathrm{in}} =|0\rangle \langle 0|$ and
$\rho_{\mathrm{f}} =|1\rangle\langle 1|$, respectively. Define the two
Kraus operators:
\begin{equation} \label{pr2pr}
K_1 = x_1|1\rangle\langle 0|+x_2 |0\rangle\langle 1|,
\qquad K_2 = x_3|1\rangle\langle 0|+x_4 |0\rangle\langle
1|,
\end{equation}
where $x_i\in\mathbb C$, $|x_1|^2+|x_3|^2=1$, and
$|x_2|^2+|x_4|^2=1$. The operators (\ref{pr2pr}) satisfy
the normalization condition (\ref{eq:k_condi}) and define
the Kraus map\footnote[3]{Abbreviation ptp of {\it pure to
pure} is used to indicate that the corresponding Kraus map
$\Phi_{\rm ptp}$ transforms a pure state into a pure
state.}
\begin{equation} \label{pr2pr-map}
\Phi_{\mathrm{ptp}}[ \rho_{\mathrm{in}} ] = \sum_{i=1}^2 K_i
\rho_{\mathrm{in}} K_i^{\dagger} = \rho_{\mathrm{f}} ,
\end{equation}
which transforms the initial state $\rho_{\mathrm{in}}$ into the final
state $\rho_{\mathrm{f}}$.

If the Kraus operators $K_1$ and $K_2$ of (\ref{pr2pr})
are linearly dependent, $K_1 = z K_2$ ($z \in
\mathbb{C})$, then $\Phi_{\mathrm{ptp}}$ is a unitary map.
However, if the Kraus operators $K_1$ and $K_2$ are
linearly independent, the corresponding Kraus map
$\Phi_{\mathrm{ptp}}$ represents \emph{non-unitary}
evolution that steers a two-level open quantum system
between two pure states. In this case, the choice of the
parameters $x_i$ in (\ref{pr2pr}) is arbitrary up to the
conditions of normalization and linear independence, and
therefore there exist infinitely many pairs of Kraus
operators $\{ K_1 , K_2 \}$ which define different Kraus
maps with the same property~(\ref{pr2pr-map}). Note also,
that Kraus maps $\Phi_{\mathrm{ptp}}$ of (\ref{pr2pr-map})
are one-to-one maps and differ from the all-to-one map
$\Phi$ defined by (\ref{eq:Phi-tot-2}). This emphasizes
the existence not only of a multitude of different
operator-sum representations of the same map, but of
qualitatively different Kraus maps, all of which are
capable of moving the open quantum system between the same
pair of states by non-unitary dynamics.

The influence of the environment on an open quantum system is
typically viewed as hindering unitary control pathways which would be
otherwise effective for the closed system. However, the possibility of
transforming pure states into pure states via non-unitary dynamics
reveals a plethora of control pathways for open quantum systems. The
existence of a multitude of non-unitary control pathways implies
flexibility and possibly control robustness in the sense that if some
transitions are blocked due to dynamical restrictions, other pathways
may still allow the controls to move the dynamics forward. The
existence of non-unitary controls, which nevertheless maintain
coherence of the initial state, may be useful for quantum information
applications in which the loss of coherence is a serious impediment.

\section{Conditions for dynamic state controllability of
open quantum systems with Kraus-map evolution}
\label{sec:DSC}

An important question yet to be fully resolved is DSC of open quantum
systems. In order to study the problem of DSC one needs to specify the
dynamical capabilities, i.e., the set of available controls. While for
a closed quantum system with unitary dynamics all available controls
are coherent, the Kraus-map dynamics of an open system can be induced
by both coherent and incoherent controls (the former act only through
the Hamiltonian part of the dynamics, while  the latter include
interactions with other quantum systems and measurements). Let
$\mathcal{C}$ be a set of all available finite-time controls, which
may include coherent electromagnetic fields, tunable distribution
functions of various environments, measurements, etc. Each particular
configuration of controls, $c(t) \in \mathcal{C}$, induces the
corresponding time evolution of the system through the Kraus map
$\Phi_{c,t}$ that transforms an initial state $\rho(0)$ into the state
$\rho(t) = \Phi_{c,t}[\rho(0)]$ at time $t$. Based on these
considerations, we introduce the following definition of open-system
DSC:
\begin{definition} \label{def:KM-DSC}
An open quantum system with Kraus-map evolution is dynamically
controllable in the set $\mathcal{S}_{\mathrm{D}}$ of states if for
any pair of states $\rho_1 \in \mathcal{S}_{\mathrm{D}}$ and $\rho_2
\in \mathcal{S}_{\mathrm{D}}$, there exists a configuration of
controls $c(t) \in \mathcal{C}$ and a finite time $T$, such that the
resulting Kraus map $\Phi_{c,T}$ transforms $\rho_1$ into $\rho_2$:
$\rho_2 = \Phi_{c,T} [\rho_1 ]$.
\end{definition}
We can also generalize the definition of DSC by considering the
asymptotic evolution, $t \to \infty$; the corresponding state is
defined as $\rho(\infty) = \lim\limits_{t \to \infty}
\Phi_{c,t}[\rho(0)]$ (if the limit exists). The system is
asymptotically controllable if the case of $t \to \infty$ is included
in Definition~\ref{def:KM-DSC}.

Complete DSC will be achieved under the Kraus-map dynamics if
$\mathcal{S}_{\mathrm{D}}$ coincides with  $\mathcal{S}_{\mathrm{K}} =
\mathcal{D}(\mathcal{H})$, i.e., it includes all density operators
$\rho$ on the Hilbert space $\mathcal{H}$ of the system. Similar to
EOC of closed quantum  systems, we can also define Kraus-map
controllability (KMC) of open systems:
\begin{definition} \label{def:KMC}
An open quantum system with Kraus-map evolution is Kraus-map
controllable if the set $\mathcal{C}$ of all available controls
generates all maps $\Phi$ of the form~(\ref{eq:Kraus-map}) from the
identity map $I$.
\end{definition}
Definitions~\ref{def:KM-DSC} and \ref{def:KMC} allow us to
formulate another corollary of Theorem~\ref{t1}:
\begin{corollary} \label{c2}
KMC is sufficient for complete DSC of a finite-dimensional open
quantum system.
\end{corollary}
If any Kraus map can be generated by available controls, then,
according to Theorem~\ref{t1} (or, more specifically,
Corollary~\ref{t1w}), any state-to-state transition can be enacted. In
general, complete DSC is weaker than KMC, since the former can be
achieved even if the set of available controls generates only a subset
of all possible Kraus maps. For example, generating all Kraus maps of
the form~(\ref{eq:Phi-tot-2}) is sufficient for making possible all
state-to-state transformations. These maps form only a subset of all
Kraus maps, but nevertheless the ability to generate all maps in this
subset using some set of controls implies complete DSC.

There can be various dynamical methods to engineer
arbitrary finite-time Kraus-map dynamics of open quantum
systems. One method relies on the ability to coherently
control both the system and environment. Let the system
under control be characterized by a Hilbert space ${\cal
H}_1$ of dimension $N$. An arbitrary Kraus map
$\Phi[\rho]=\sum_{i=1}^n K_i\rho K_i^\dagger$ (where $n$
can be chosen such that $n\le N^2$) in the space of states
${\cal D}({\cal H}_1)$ can be realized by coupling the
system to an ancilla (which serves as an effective
environment), characterized by Hilbert space ${\cal H}_2$
of dimension $n$, and generating a unitary evolution
operator $U$ acting in the Hilbert space of the total
system ${\cal H}={\cal H}_1\otimes{\cal H}_2$ as
follows~\cite{Preskill}. Choose in ${\cal H}_2$ a vector
$|0\rangle$ and an orthonormal basis $|e_i\rangle$,
$i=1,\dots,n$. For any $|\psi\rangle\in {\cal H}_1$ let $U
(|\psi\rangle\otimes|0\rangle)=\sum\limits_{i=1}^n
K_i|\psi\rangle\otimes |e_i\rangle$. Then, such operator
can be extended to a unitary operator in $\cal H$ and for
any $\rho\in {\cal D}({\cal H}_1)$ one has
$\Phi[\rho]={\rm Tr}_{{\cal H}_2}\,\{ U(\rho\otimes
|0\rangle\langle 0|)U^\dagger\}$. Therefore the ability to
dynamically create, for example via coherent control, an
arbitrary unitary evolution of the system and ancilla
allows for generating arbitrary Kraus maps of the
controlled system. In terms of controllability of
finite-dimensional systems, this means that EOC of the
system and environment taken together is sufficient for
KMC and, according to Corollary~\ref{c2}, for complete DSC
of the system as well. Since in practice one may not have
full control over the environment, Lloyd and Viola
\cite{LV02} proposed another method of Kraus-map
engineering, based on a combination of coherent controls
and measurements. They have shown that the ability to
perform a single simple measurement on the system,
together with the ability to apply coherent control to
feed back the measurement results, allows for enacting an
arbitrary finite-time Kraus-map evolution of the
form~(\ref{eq:Kraus-map}). This procedure determines
another set of controls that is sufficient for KMC and,
according to Corollary~\ref{c2}, for complete DSC of a
finite-dimensional open quantum system.

\section{Conclusions}
\label{sec:disc}

This paper establishes and illustrates complete KSC of
finite-dimensional open quantum systems with Kraus-map
dynamics. The main theoretical result of the paper is that
for any target state $\rho_{\mathrm{f}}$ on the Hilbert
space of the system, there exists a Kraus map that
transforms \emph{all} initial states into
$\rho_{\mathrm{f}}$. The possibility of designing control
operations which steer all initial states of the system to
a given target state is available due to the use of
non-unitary dynamics and is in principle unattainable in
the framework of unitary control. Non-unitary controls
also allow for transformations between specific pure and
mixed states and vice versa. Moreover, there exist
different Kraus maps which perform a desired
state-to-state transition. For example, the transition
between a pair of pure states can be performed via three
qualitatively different families of Kraus maps: (i)
unitary transformations, (ii) all-to-one non-unitary maps,
and (iii) one-to-one non-unitary maps.

General definitions of DSC and KMC were introduced for
finite-dimensional open quantum systems, leading to the
result that KMC is a sufficient condition for complete
DSC. Thus, if the available control tools make possible
enacting any Kraus map, then any initial state on the
Hilbert space of the system can be transformed into any
final state. Combining this result with prior findings
\cite{LV02} on Kraus-map engineering determines two
specific sets of control operations (i.e., the two methods
described in the last paragraph of section~\ref{sec:DSC})
which are sufficient for the system to be dynamically
controllable in the set of \emph{all} states on the
Hilbert space. Moreover, any dynamical control which can
produce a Kraus map of Theorem~\ref{t1} by one of these
approaches, will be robust to variations in the initial
state. Important problems for future research could be to
establish the necessary conditions for dynamical
controllability of specific open quantum systems and to
analyze robustness of the dynamical control to
imperfections and environmental effects during the
evolution.

\ack
The authors thank Ian Walmsley and Robert Kosut for helpful
discussions.  This work was supported by the Department of
Energy. Rong Wu was supported by a fellowship from the Program in
Plasma Science and Technology.

\end{document}